\documentclass[12pt,a4paper]{article}
\usepackage{color,amsmath,amsfonts,graphics,epsfig,amssymb}
\renewcommand{\baselinestretch}{1.1}
\makeatletter
 \long\def\@makecaption#1#2{%
   \vskip\abovecaptionskip
     \renewcommand\baselinestretch{0.95}
     \normalfont
     \setbox\@tempboxa\hbox{#1. #2}
     \ifdim \wd\@tempboxa < \textwidth
       \centerline{\box\@tempboxa}
     \else
       #1. #2
     \fi
   \vskip\belowcaptionskip}
 \makeatother
\begin{document}
\title{\large \bf Estimate of the fraction of primary photons in the
cosmic-ray flux at energies $\bf{\sim 10^{17}}$~eV from the EAS-MSU
experiment data
}
\author{\large Yu.A.~Fomin$^{a}$, N.N.~Kalmykov$^{a}$, G.V.~Kulikov$^{a}$,
V.P.~Sulakov$^{a}$  \\
and
S.V.~Troitsky$^{b}$
\\[12pt]
\small \it
$^a$~D.V.~Skobeltsyn Institute of Nuclear Physics, \\
\small \it
M.V.~Lomonosov Moscow State University, Moscow 119991, Russia
\\
\small \it
$^b$~Institute  for Nuclear Research of the
Russian Academy of Sciences, Moscow 117312, Russia
}
\date{}
\maketitle
\begin{abstract}We reanalyze archival EAS-MSU data in order to search for
events with anomalously low content of muons with energies $E_{\mu}>10$~GeV
in extensive air showers with number of particles $N_{e} \gtrsim 2\times
10^{7}$. We confirm the first evidence for nonzero flux of primary cosmic
gamma rays at energies $E\sim 10^{17}$~eV. The estimated fraction of
primary gamma rays in the flux of cosmic particles with energies $E
\gtrsim 5.4 \times 10^{16}$~eV is $\epsilon_{\gamma}=\left(
0.43^{+0.12}_{-0.11} \right)$\%, which corresponds to the intensity of
$I_{\gamma}=\left(1.2^{+0.4}_{-0.3} \right)\times 10^{-16}~{\rm
cm}^{-2}{\rm s}^{-1}{\rm sr}^{-1}$. The study of arrival directions does
not favour any particular mechanism of the origin of the photon-like
events.
\end{abstract}

\section{Introduction}
\label{sec:intro}

The study of the primary mass composition of ultra-high-energy (UHE)
cosmic rays (CR) is one of the topical problems of astroparticle
physics because these experimental results are of crucial importance for
understanding the theory of both cosmic-ray generation in their sources
and their subsequent propagation to the Earth. Low UHECR intensity
makes their study by direct methods impossible, so that the only available
method is the study of extensive air showers (EAS).

The dominant part of EAS is caused by primary nuclei (from protons to
iron), however, there is a considerable interest to possible presence of
very different particles, e.g.\ UHE gamma rays, among them.
First works on the subject appeared already a half-century ago (see
e.g.\ Ref.~\cite{KhristiansenBook}) but
definitive quantitative results are still missing (cf.\ a review
\cite{RisseReview} and references therein).
Indeed, the highest-energy cosmic photons firmly
detected had the energy of $\sim 50$~TeV \cite{MILAGRO-90TeV}. The
searches for gamma rays in the energy ranges $3\times 10^{14}$~eV$\lesssim
E \lesssim 5 \times 10^{16}$~eV (the EAS-TOP~\cite{EAS-TOP},
CASA-MIA~\cite{CASA-MIA} and KASCADE~\cite{KASCADE} experiments) as well
as at $E \gtrsim 10^{18}$~eV (the Haverah Park \cite{Haverah}, AGASA
\cite{AGASA, RisseHomola, AGYak}, Yakutsk \cite{Yak1, Yak2}, Pierre Auger
\cite{PAO1, PAO2} and Telescope Array \cite{TA} experiments) did not
find any signal and resulted in upper limits on the photon flux only. A
few claims of the experimental detection of $10^{14}$~eV$\lesssim
E \lesssim 10^{17}$~eV photons (Mt.~Chacaltaya~\cite{Chac}, Tien
Shan~\cite{Tien}, Yakutsk~\cite{YakClaim} and Lodz~\cite{Gawin}) had low
statistical significance. At the same time, a certain flux of UHE photons
is predicted in many models of both conventional and ``new'' physics. In
particular, the flux of secondary photons from interactions of
extreme-energy particles with cosmic background radiation, the so-called
Greizen-Zatsepin-Kuzmin (GZK) photons, may serve as a tool to distinguish
various models of cosmic rays at energies $\gtrsim 5\times 10^{19}$~eV
because the photon flux is very sensitive to the primary composition at
these energies: predominantly light composition at GZK energies results in
a much higher flux of secondary photons. Given the present contradictory
situation with the mass composition at UHE (see e.g.\ Ref.~\cite{Unger}
for a detailed review and Ref.~\cite{ST-UFN} for a brief update), searches
for GZK photons are now considered very important. Also, a significant
contribution to the UHE gamma-ray flux is predicted in particular top-down
mechanisms of CR origin (\cite{BhSigl} and references therein), in
particle-physics models with Lorentz-invariance violation
\cite{Lorentz-violation} and in models with axion-photon mixing
\cite{axion}.

One of the most promising approaches to the search of primary gamma rays
is the study of the EAS muon component. The number of muons in a gamma-ray
induced EAS is an order of magnitude smaller than in a usual hadronic
shower. Therefore, one may hope to find photon showers by selecting those
which have unusually low muon content.

In the present work, we study the muon content of showers with the
estimated number of particles  $N_{e}>2\times 10^{7}$
and zenith angles $\theta<30^{\circ}$ detected by the EAS-MSU array
\cite{EAS-MSU} in 1982 -- 1990. We demonstrate that the number of muonless
events exceeds significantly the background expected from random
fluctuations in the development of showers caused by primary hadrons. This
result may be interpreted as an indication to the presence of gamma rays
in the primary cosmic radiation with energies of order $10^{17}$~eV which
confirms and strengthens the first evidence for UHE cosmic photons
\cite{Khorkhe}.

The rest of the paper is organized as follows. In
Sec.~\ref{sec:experiment}, we briefly review the experimental setup
(Sec.~\ref{sec:installation}), then discuss the data set we study, and
muonless events in particular (Sec.~\ref{sec:data}). Sec.~\ref{sec:flux}
is devoted to the estimate of the number of background muonless events for
hadronic showers (Sec.~\ref{sec:MC}) and to the derivation of the
estimated photon flux under the assumption that all muonless events not
accounted for by the hadronic background are caused by primary gamma rays
(Sec.~\ref{sec:estimate}). Possible systematic errors in the determination
of the flux are discussed in Sec.~\ref{sec:syst}. In
Sec.~\ref{sec:arr-dir}, we present a detailed study of the distribution of
the arrival directions of muonless events on the celestial sphere and test
various models of the origin of primary photons. We put our results in the
context of the present-day \textit{state of the art} and briefly conclude
in Sec.~\ref{sec:concl}.

\section{Experiment and data}
\label{sec:experiment}
\subsection{The EAS-MSU array}
\label{sec:installation}
\begin{figure}
\centering \includegraphics[width=0.75\columnwidth]{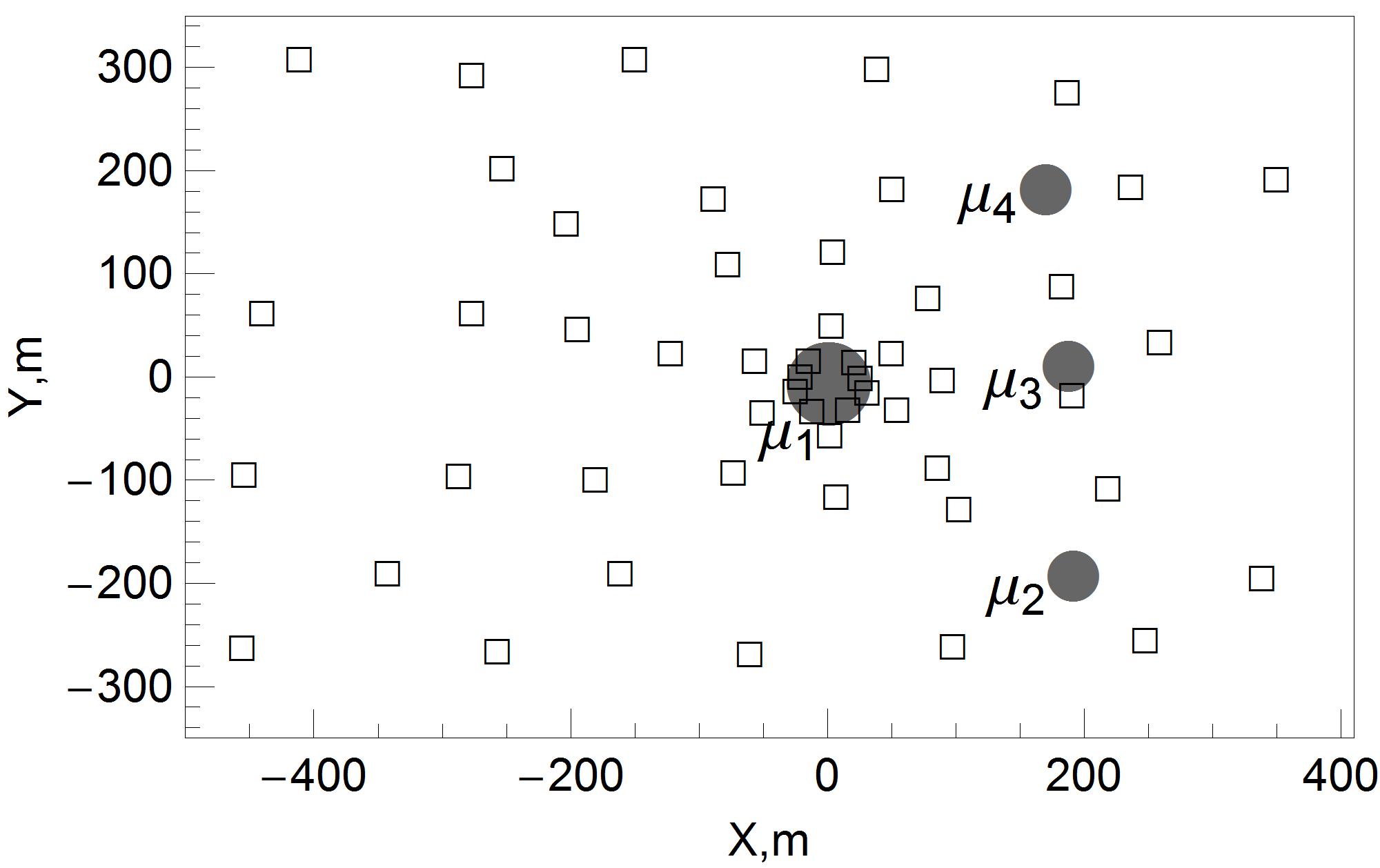}
\caption{
\label{fig:installation}
The EAS-MSU array setup. Muon detectors (circles) are denoted by
$\mu_{i}$, $i=1,\dots,4$; surface detector stations are represented by
squares. }
\end{figure}

The description of the EAS-MSU array is given in~\cite{EAS-MSU}. The
array had the area of 0.5~km$^2$ and contained 77
charged particle density detectors (consisted of the Geiger-Mueller
counters) for determination of the EAS size $N_e$ employing the empirical
lateral distribution function~\cite{muLDF} and 30 scintillator detectors
which measured particle arrival times necessary for determination of the
EAS arrival direction. In addition to the surface detectors which recorded
mostly electron-photon component of an EAS, the array included also four
underground muon detectors, also consisted of Geiger-Mueller counters,
located at the depth of 40~meters of water equivalent. These detectors
recorded muons with energies above 10~GeV. A muon detector with the area
of 36.4~m$^{2}$ was located at the center of the array while other three
stations had the area of 18.2~m$^{2}$ and were located at the distances
between 150~m and 300~m from the center (see Fig.~\ref{fig:installation}).
To select the sample of showers with the number of particles
$N_{e}>2\times 10^{7}$ which we use in this work, 22 scintillator
detectors, each of the area of 0.5~m$^{2}$,  were used. The
scintillator detector threshold was set at the level of $1/3$ of a
relativistic particle. The temporal resolution was $\sim 5$~ns. The 22
stations form 13 systems of 4-fold coincidences between counters located
at the vertices of tetragons with sides between 150~m and 300~m which
allowed one to select efficiently the showers on the full array
area. The scintillator detectors were located at the same points as the
Geiger-Mueller counters. The master criterion was determined by the
firing, in the time gate of $\sim 6~\mu$s, of at least one of the 4-fold
coincidence systems.

To reduce the number of sub-threshold events which still satisfy the
master conditions, the express-analysis of the number of fired
Geiger-Mueller counters was invoked. In each case, it was required that at
least 4 of 22 counters in the selection system recorded the density
exceeding 1 particle per square meter. With these selection criteria
implemented, the probability of detection of a shower with $N_{e}>2\times
10^{7}$ falling to any place of the array was not less than 95\%. The
position of the shower axis was determined with the precision of $\sim
10$~m. The precision of determination of the arrival direction was
$\sim 3^{\circ}$. The number of particles in the shower was determined with
the accuracy $\sim (15-20)\%$.

\subsection{The data set and muonless events}
\label{sec:data}
The presence of muon detectors in the EAS-MSU array allows to
search for primary gamma rays. The method is based on the fact that, for
$N_{e} \gtrsim 10^{7}$ and for an hadronic primary, it is highly
unprobable to have zero muons in the central, 36.4~m$^{2}$, detector if the
shower axis is within $\sim 240$~m from it. At the same time, these
muonless events are fully consistent with the conjecture of primary gamma
rays. The total number of events with $N_{e}\ge 2 \times 10^{7}$ in the
data set is 1679; of them 48 are muonless.

\begin{figure}
\centering \includegraphics[width=0.65\columnwidth]{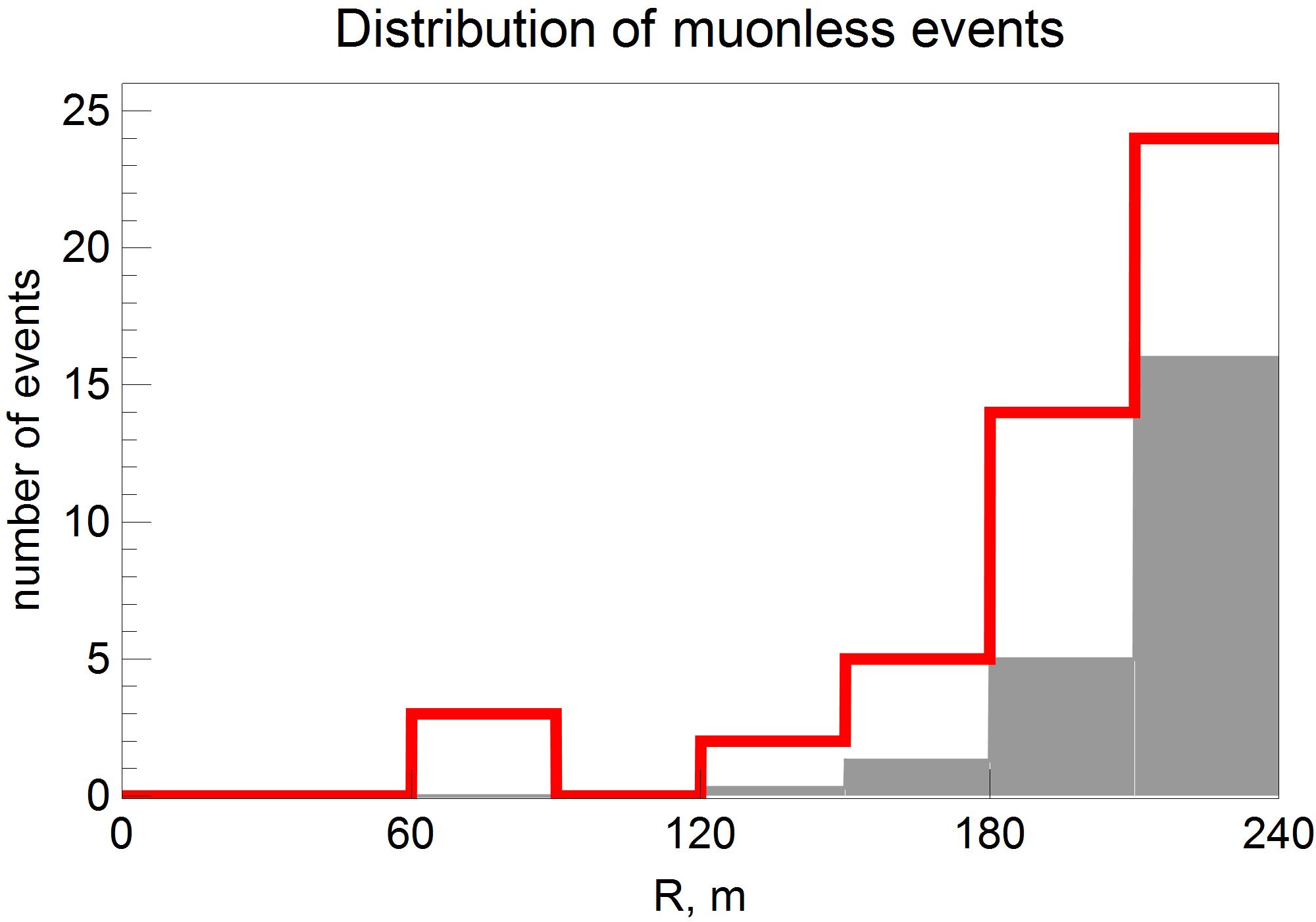}
\caption{
\label{fig:Rdistribution}
The distribution of muonless events in the distance $R$ between the shower
axis and the muon detector. Line: data; shadow: expectation for
hadronic primaries. }
\end{figure}

Fig.~\ref{fig:Rdistribution}
presents the distribution of muonless events in the distance $R$ between
the shower axis and the muon detector. Most of the muonless events
correspond naturally to large $R$; however, there are a certain number of
events close to the axis which are very difficult to explain by random
fluctuations of the hadronic background. One should note that the real
number of muonless events is larger than observed because of the non-EAS
background which results in firing of each counter in the central muon
detector with average frequency of 4.6~Hz. In three other muon detectors,
the frequency of random firing was 2 to 3 times higher, and in this work,
we use only the data of the central detector. It consisted
of 1104 counters. For the time of EAS detection $\sim 15~\mu$s, one
expects 0.076 random firings. Therefore we assume that the probability of
absence of the random firing was 0.93.

\begin{figure}
\centering \includegraphics[width=0.65\columnwidth]{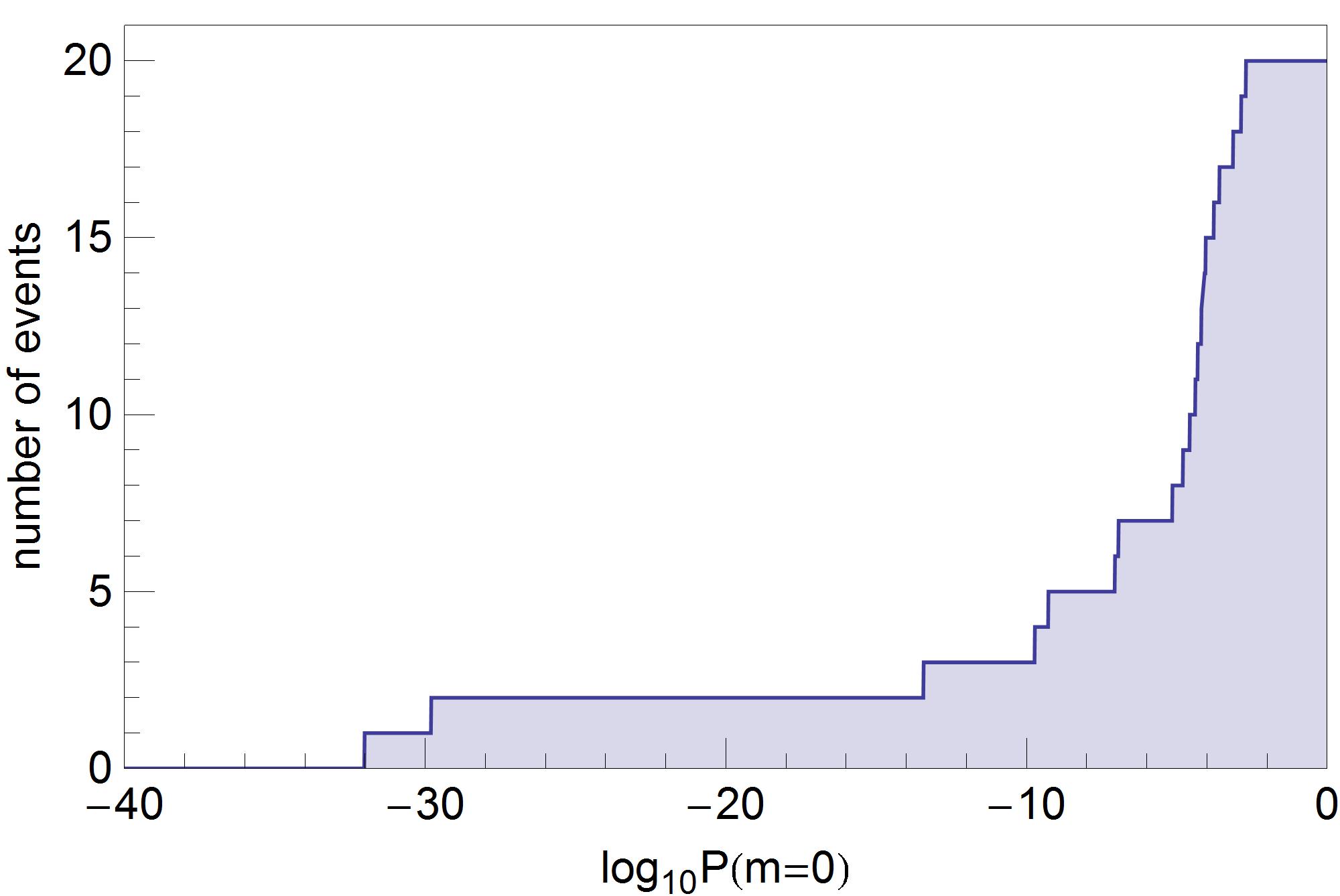}
\caption{
\label{fig:P-Poisson}
Cumulative distribution of muonless events in the LDF-based Poisson
probability $P(m=0)$. }
\end{figure}

To obtain a very rough estimate of the probability to have a muonless
hadronic event, one may start with the (experimentally known) mean muon
lateral distribution function \cite{muLDF} and estimate the expected muon
density $\rho_{\mu}(N_{e},R)$ for a given core distance $R$. Then, by
making use of the Poisson distribution, one may calculate the probability
$P(m=0)$ to have no muons in the detector at this distance. In
Fig.~\ref{fig:P-Poisson},
the distribution of $m=0$ events in $P(m=0)$ is shown. The tail at low
$P(m=0)$ indicates that there might be a problem in explaining the
observed number of muonless events within the standard model of the shower
development.

\section{Estimates of the gamma-ray flux}
\label{sec:flux}
To quantify the observed discrepancy more precisely, we performed
Monte-Carlo simulations of proton-induced showers and compared the number
of muonless events in data and in simulations.
\subsection{Modelling of artificial showers}
\label{sec:MC}
\begin{figure}
\centering \includegraphics[width=0.65\columnwidth]{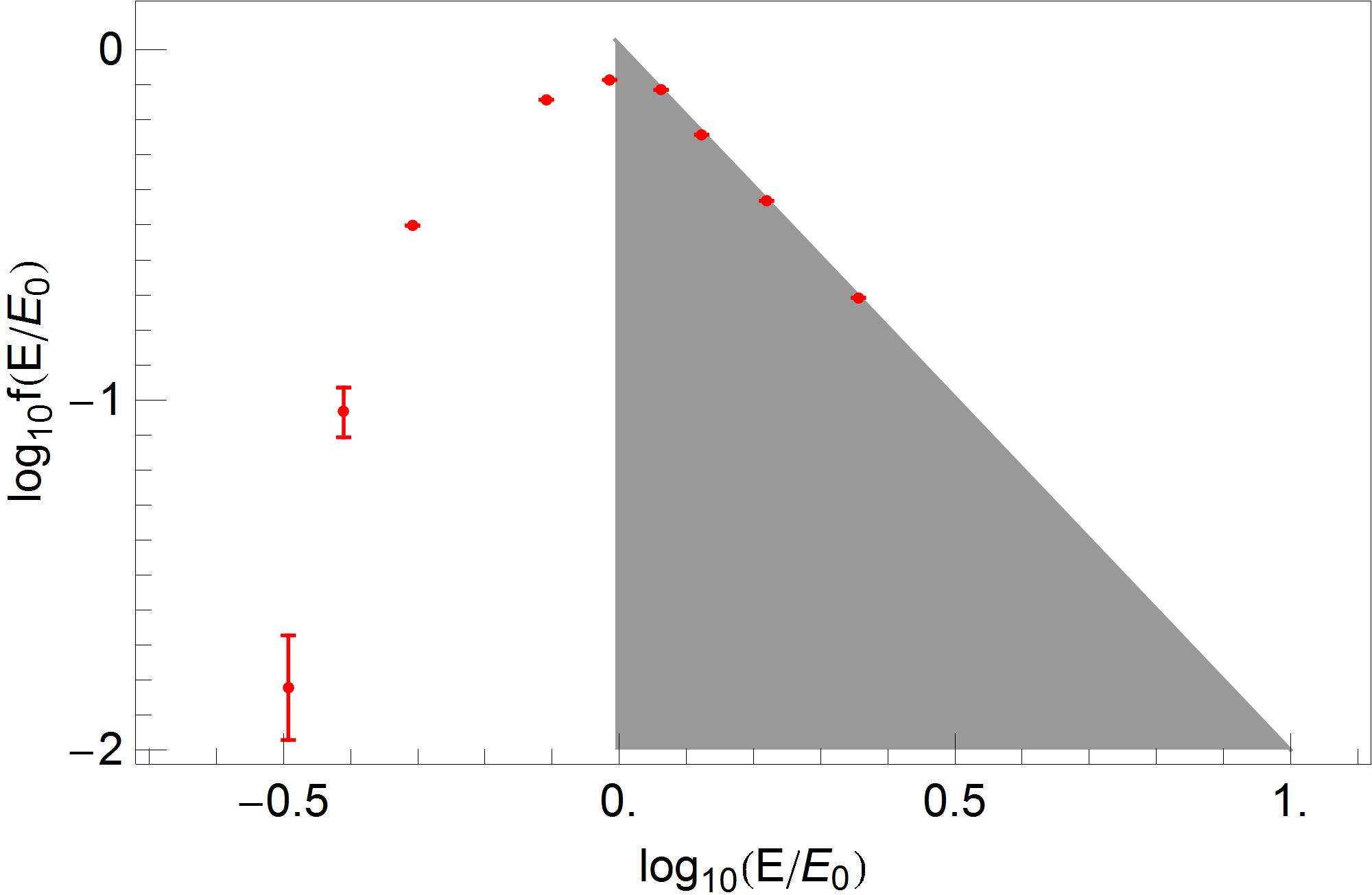}
\caption{
\label{fig:E-Ne-fluct}
Contribution of various primary energies to the proton showers with $N_{e}
\ge 2\times 10^{7}$. Points: results of the simulation; shadow: naive
 estimate without the account of fluctuations. $E_{0}\approx 10^{17}$~eV.
}
\end{figure}

For the shower simulations, we used the AIRES v.~2.6.0 \cite{AIRES}
simulation code, whose choice was determined primarily by its speed. We
used the high-energy hadronic interaction model QGSJET-01 \cite{QGSJET}.
The primary protons were thrown with zenith angles $0^{\circ} \le \theta
\le 30^{\circ}$ and with energies between $3\times 10^{16}$~eV and
$2\times 10^{17}$~eV, assuming the integral spectral index of 2.0. Without
the account of fluctuations, the energy of a $N_{e}=2\times 10^{7}$
proton shower would be equal  to $E\sim 10^{17}$~eV; however, the
fluctuations reduce this value. For the study, showers with $N_{e} \ge
2\times 10^{7}$ have been selected; Fig.~\ref{fig:E-Ne-fluct}
gives the distribution of the primary energies of the selected artificial
showers. In this way, the total amount of 15000 artificial showers were
simulated.

\subsection{Estimate of the fraction and flux of gamma rays}
\label{sec:estimate}
The general assumption behind our estimate of the gamma-ray flux is that
all muonless events, not accounted for by fluctuations of hadronic
showers, are caused by primary gamma rays. Therefore, the central moment
of the estimate is the calculation of the expected number of background
muonless events from the simulated proton-induced showers.

The probability
of a zero muon detector reading, $m=0$, was estimated under assumption
(see Ref.~\cite{assumption} for its motivation) that the fluctuations of
muon density in EAS may be represented as a superposition of
(a)~fluctuations of muon density at a given distance from the shower axis,
determined purely by the EAS development, and (b)~the Poisson fluctuations
of the number of particles which hit the detector station. In this
approach, the probability $P(\Delta R, S,m)$ to have $m$ muons in the
detector of area $S$ located in the ring $\Delta R$ (distance between
$R$ and $R+\Delta R$ from the shower axis), is given by
$$
P(\Delta R, S,m) =\int\! P_{\rm EAS} (\Delta R,M) \cdot P_{\rm P}(M,S,m) \,
dM,
$$
where $P_{\rm
EAS} (\Delta R,M)$ is the function of the muon number density distribution
in the ring as determined by the shower development and $P_{\rm P}(M,S,m)$
is the Poisson probability to record exactly $m$ muons in a detector of the
area $S$ for the total number $M$ of muons in the ring.

Suppose that a shower axis came within the ring $\Delta
R_{k}=R_{k+1}-R_{k}$ from the muon detector. Then the muon density in the
ring is determined as
$$
\rho_{\mu}(\Delta R_{k}, i) =
\frac{N_{\mu }(\Delta R_{k}, i)}{\left(\pi \left(
R_{k+1}^{2}-R_{k}^{2}\right) \right)}
,
$$
where $N_{\mu }(\Delta R_{k}, i)$ is the number of muons in this ring and
$i=1, \dots n_{\rm tot}$ is the number of the selected artificial showers.
Then the probability of a zero detector reading in the $\Delta R_{k}$ ring
is
$$
P(m=0,\Delta R_{k})=\frac{1}{n_{\rm tot}} \sum\limits_{i}^{n_{\rm tot}}
0.93 \exp\left(-S \cos\theta\, \rho_{\mu}(\Delta R_{k},i)   \right),
$$
where $\theta$ is the zenith angle of the shower.

The total probability of a $m=0$ event is
$$
P_{\rm tot}(m=0)=\sum\limits_{k=1}^{k_{\rm max}}P(m=0,\Delta R_{k})
\frac{R_{k+1}^{2}-R_{k}^{2}}{R_{k_{\rm max}}^{2}},
$$
where $k_{\rm max}$ gives the total number of rings considered and the
last factor accounts for the probability for the shower axis to hit the
$\Delta R_{k}$ ring. The results of the calculation of probability to
observe a muonless event are given, for various distances from the shower
axis, in Table~\ref{tab:prob}
\begin{table}
\centering
\begin{tabular}{ccccccc}
\hline
\hline
$\Delta R$, m && Observed number of && $P(m=0, \Delta R)$ && Expected
number of\\
&&muonless events && && muonless events\\
\hline
60--90 && 3 && $1.8 \times 10^{-7}$ && $3\times 10^{-4}$\\
90--120 && 0 && $7.7 \times 10^{-6}$ && 0.013\\
120--150 && 2 && $2.15 \times 10^{-4}$ &&0.36 \\
150--180 && 5 && $8.0 \times 10^{-4}$ &&1.3 \\
180--210 && 14 && $3.2 \times 10^{-3}$ &&5.4 \\
210--240 && 24 && $9.8 \times 10^{-3}$ &&16.4 \\
\hline
0--240 && 48 && $1.4 \times 10^{-2}$ && 23.5\\
\hline
\hline
\end{tabular}
\caption{\label{tab:prob}
Observed vs.\ expected number of muonless events for various distances
between the detector and the shower axis. See text for notations. }
\end{table}
together with the number of observed and predicted muonless events in our
sample of 1679 showers.

The total probability to have a muonless proton-induced event within 240~m
between the detector and the shower axis is $1.4\times 10^{-2}$ which
corresponds to $\approx 23$ expected muonless events in the sample, to
compare with 48 observed. As expected, the dominant part of the background
muonless events should appear in two outer rings we considered, the same
being true also for the observed events. However, the total number of the
observed events is almost twice the expected one. This allows one to
estimate, based on the Poisson distribution, the number $S$ of signal
photon-like events in the sample as
$
S=25.2^{+7.2}_{-6.6},
$
which transforms into the fraction
$\epsilon_{1}=\left(1.50^{+0.43}_{-0.39} \right)\%$ of anomalous muonless
events in the sample with $N_{e} \ge 2\times 10^{7}$ and $\theta \le
30^{\circ}$.

We want to identify the anomalous muonless showers with showers initiated
by primary photons. To determine the fraction of these events in the
energy spectrum of cosmic rays, one needs to take into account the
difference in the development of showers caused by photons and protons of
the same energy. The gamma-ray showers develop slower in the atmosphere
and arrive younger to the surface level (the vertical atmospheric
depth for EAS-MSU is 1025~g$\cdot$cm$^{-2}$). On average, for the primary
energies $\sim 10^{17}$~eV, the number of particles in a gamma-ray shower
detected by the EAS-MSU experiment should be $\approx 1.86$ times larger
compared to the proton shower. The cut in $N_{e}$ we use thus corresponds,
on average, to the gamma-ray energy of $5.4 \times 10^{16}$~eV. Knowing
the total cosmic-ray flux measured by the EAS-MSU
array~\cite{EAS-MSUflux}, we determine the main result of the present
work: the photon fraction,
$$
\epsilon_{\gamma}=\left(0.43 ^{+0.12}_{-0.11} \right)\% ~~~\mbox{for}~E
\gtrsim 5.4 \times 10^{16}~\mbox{eV},
$$
and the photon flux intensity,
\begin{equation}
I_{\gamma}=\left( 1.2^{+0.4}_{-0.3}  \right)\times 10^{-16}~
{\rm cm}^{-2}\,{\rm s}^{-1}\,{\rm sr}^{-1}
~~~\mbox{for}~E
\gtrsim 5.4 \times 10^{16}~\mbox{eV}.
\label{eq:flux}
\end{equation}

\subsection{Estimate of systematic uncertainties}
\label{sec:syst}
The systematic uncertainty of our result, within the method we use, is
related to the estimate of the number of background muonless events from
hadronic showers.

\textbf{Hadronic interaction models.}
The largest uncertainty comes from the variety of
models of shower development which predict different values of muon number
in EAS. Furthermore, this difference is sensitive to the muon threshold
energy, which is 10~GeV in our case. The change of the mean expected muon
density in EAS by $\pm 10\%$ would result in the change of the number
of background muonless showers in the sample by $\pm 4$. The results we
quote are based on the QGSJET-01 model \cite{QGSJET} which gives a good
description of the LHC and Pierre Auger Observatory measurements of the
high-energy hadronic cross section (cf.\ Fig.~5 of Ref.~\cite{ST-UFN}) and
of the LHC multiplicity distributions (see e.g.\ Ref.~\cite{Q01-mult});
the
choice of the model was also motivated by its computational efficiency.
The amount of model-to-model variations of the number of $>10$~GeV muons
in EAS may be estimated from Ref.~\cite{EngelTalkUHECR} and from our own
simulations. The effect of the change of the interaction model on our
results is summarized in Table~\ref{tab:models}.
\begin{table}
\centering
\begin{tabular}{ccccccc}
\hline
\hline
Model && $\displaystyle \frac{N_{\mu}}{N_{\mu}(\mbox{QGSJET-01})}$ &&
Expected number of && Excess number of \\
&&  &&muonless events && muonless events\\
\hline
SIBYLL~2.1 \cite{SIBYLL}    && 0.70 && 38.0 && 10.7\\
QGSJET~II-03 \cite{QGSJET2} && 0.89 && 27.6 && 21.0\\
\hline
QGSJET-01~\cite{QGSJET}     && 1.00 && 23.5 && 25.2\\
\hline
EPOS~1.99~\cite{EPOS}       && 1.03 && 21.9 && 26.8\\
experiment~\cite{PAOmuons}  && 1.33 && 10.9 && 37.8\\
\hline
\hline
\end{tabular}
\caption{\label{tab:models}
Effect of the choice of hadronic interaction models on the result. The
last line corresponds to experimental results on the muon content of EAS. }
\end{table}
We shall note that, according to experimental data on EAS development, all
hadronic-interaction models currently in use underestimate the number of
muons in a shower significantly. In particular, several independent
indirect analyses of the Pierre Auger Observatory data
indicate~\cite{PAOmuons} that the real number of muons is approximately
1.5 times larger than predicted by the QGSJET~II-03 model. This number is
used in Table~\ref{tab:models} and for the estimate of the systematic
error; a similar result was obtained with the help of muon detectors of
the Yakutsk EAS array~\cite{Yakutsk-mu, Yakutsk-mu1}. The systematic error
in the resulting gamma-ray flux due to the uncertainty of hadronic models
is $\pm 50\%$, with the upper value favoured by the experimental data.

\textbf{Primary composition.}
Assumption of the purely proton composition gives a conservative (i.e.\
large) estimate of the expected background of the muonless events because
primary heavier nuclei produce more muons in EAS. For primary iron, the
corresponding number of muons is larger by a factor of $\sim 2.5$ which
shifts the expected background downwards to zero. This would change our
fraction and flux estimates by $+90\%$.

\textbf{Large fluctuations.}
Since no model gives a perfect description of hadron-induced air showers,
and in particular there are large uncertainties in predictions of muon
number, one cannot exclude that the fluctuations of the EAS muon content
might be much larger than suggested by simulations. Among theoretical
approaches, the probability of occasional very low muon density in a
proton shower is the highest in the model of Ref.~\cite{Uchaikin} where
the energy equipartition between positive, negative and neutral
components of the cascade was postulated. As it has been shown in
Ref.~\cite{Khorkhe-diss}, in the frameworks of this model, it is possible
to obtain the probability of $\sim 1\%$ of imitation of a gamma-ray shower
by a primary proton.
However, this model is much less physically motivated as compared to those
which are currently used in simulation codes.

To summarize the discussion of systematic uncertainties, current
experimental and theoretical understanding of the EAS properties suggests
that the flux values we obtain are conservative, though they could become
lower if physically less motivated models were used for hadronic showers.

\section{Arrival directions}
\label{sec:arr-dir}
In this section, in order to find some hints about the origin of the
events we observed, we perform various searches for deviations from
isotropy in the distribution of arrival directions of the photon-like
events. All the tests are performed by comparison, by means of a certain
statistical procedure, of the real distribution of arrival directions with
the simulated one which assumes isotropy. In all cases, the result of a
test is given by the probability $P$ that the actual distribution of
events is a fluctuation of the isotropic distribution, that is for small
$P$, the isotropic distribution is excluded at the confidence level of
$(1-P)$. For tests of the global (large-scale) isotropy, we use the
Kolmogorov-Smirnov method (see e.g.\ Ref.~\cite{nr}) which compares
one-dimensional distributions of real and simulated events in some
observable (e.g., a celestial coordinate). For searches of the local
(small-scale) anisotropy, we rely on the correlation-function method which
estimates how often the number of pair coincidences of directions from two
catalogs (e.g., one of the arrival directions of cosmic rays and another
of particular astronomical objects) in simulated samples exceeds the
similar number obtained from the real data. The notion of the ``pair
coincidence'' depends on the angular distance $\Delta$ between the
directions, so the probability $P(\Delta)$ is often quoted for a certain
range of $\Delta$. The clustering properties of the sample of the
directions are estimated by the same method with both catalogs being
identical cosmic-ray lists. More details on the method may be found e.g.\
in Ref.~\cite{cut-n-pen}.

\begin{figure}
\centering
\includegraphics[width=0.85\columnwidth]{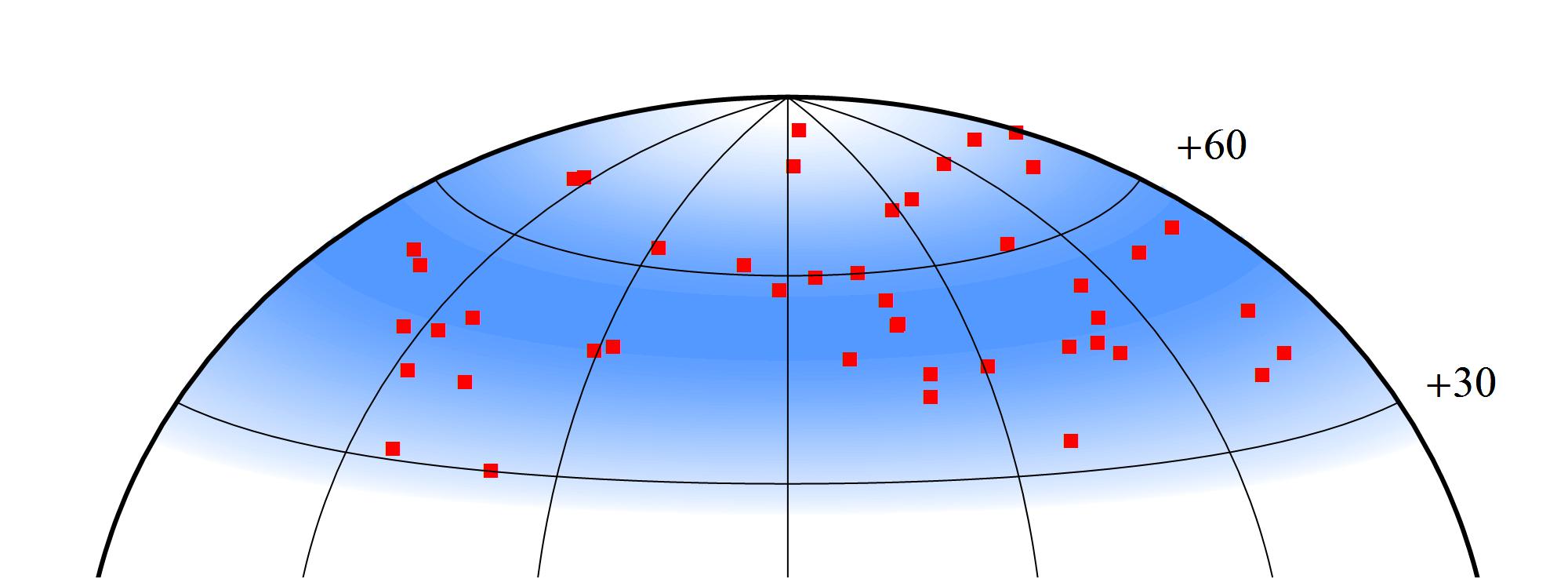}
\caption{
\label{fig:melon1}
The distribution of arrival directions of muonless events in the sky
(equatorial coordinates). Grades of shade represent the distribution
expected for the isotropic flux.}
\end{figure}

In both approaches, we need to simulate sets of arrival directions in the
assumption of the isotropic flux. These sets should take into account
the experimental selection effects. For continuously operating surface
detector arrays with the efficiency close to 100\%, the exposure is
uniform in the azimuth angle and depends on the zenith angle $\theta$ by a
purely geometric factor $\sin\theta \cos\theta$, assuming that the
incoming flux is isotropic (this is the case when one studies the
energy-limited sample of cosmic rays). However, our sample is limited by
$N_{e}$ instead of energy and, due to different age of showers coming at
different zenith angles, the exposure becomes non-geometric. Based on the
observed distribution of $\theta$, we determine the acceptance factor as
$\sim \sin\theta \cos^{9}\theta$. The distribution of events in the azimuth
angle is perfectly consistent with uniform as expected. The distribution
of the arrival directions on the sky, together with the one expected from
exposure for the isotropic flux, is shown in Fig.~\ref{fig:melon1}.
In the study of the arrival directions, we do not include 3 of 48 events
observed in 1982 for which the determination of geometry is uncertain.

\subsection{Possible scenarios for UHE photons}
Among possible mechanisms of the origin of UHE gamma rays, we consider
separately those which do not require deviations from the standard
particle-physics and astrophysical concepts (we will call these scenarios
conventional) and those which require the presence of particles and/or
interactions beyond the Standard Model of particle physics (these will be
called ``new-physics'' scenarios). Note that high-energy photons interact
with cosmic background radiation efficiently. Assuming standard physics,
the energy attenuation length for a $\sim 10^{17}$~eV photon is as low as
$\sim 35$~kpc due to efficient $e^{+}e^{-}$ pair production on the cosmic
microwave background (CMB). This means that the
observed photons were created in the Galaxy unless some new physics is
assumed.

\textbf{Conventional scenarios.}

\underline{\textit{Scenario 1. Cosmogenic photons.}}
UHE cosmic particles experience intense interactions with cosmic
background radiation. For protons with energies above $\sim 5 \times
10^{19}$`eV, these are dominated by the GZK
\cite{G, ZK} process of pion production through the $\Delta$ resonance;
for lower energies the dominant mechanism is the $e^{+}e^{-}$ pair
production. For heavier primaries at $E\sim 10^{20}$~eV,
photodisintegration effectively reduces the propagation effects to those
of protons of lower energy. The secondary particles from all these
interactions (pions, electrons and positrons) are the source of the
so-called cosmogenic photons which appear either from subsequent pion
decays or from inverse Compton scattering of $e^{\pm}$. There are a lot of
works on the GZK photons (e.g.\ Refs.~\cite{Gelmini, Sarkar}); the key
point of interest here is the possibility to use these $\sim
(10^{18}-10^{19})$~eV gamma rays as a tool to determine the composition of
the bulk of $E\sim 10^{20}$~eV cosmic rays; due to the GZK process, the
flux of the secondary photons would be much higher for super-GZK protons
than for heavy nuclei. Given the present-day uncertainty in the primary
composition at the very end of the CR spectrum, see e.g.\
Refs.~\cite{Unger, ST-UFN}, this approach attracts considerable attention,
though no sign of the GZK photons have been observed yet. The expected flux
of the GZK photons at $E \lesssim 10^{17}$~eV is far too low to explain
our result; we are not aware of a calculation of the flux at lower
energies,
nor of
the distribution of their arrival directions which however should be close
to the isotropic one.

\underline{\textit{Scenario 2. Direct photons from point sources.}}
UHE astrophysical accelerators are expected to emit energetic photons born
in interactions of charged particles with ambient matter and radiation.
The energy of accelerated particles should therefore exceed the energy of
the photons, roughly by an order of magnitude. It is presently unknown
whether the acceleration of particles up to $\sim(10^{17}-10^{18})$~eV
may happen in any single object in the Galaxy (that is, within the
propagation length of $\sim 10^{17}$~eV photons). In any case, these
objects are not expected to be numerous; we therefore expect a certain
degree of clustering of the arrival directions of photons in this
scenario. Galactic TeV gamma-ray sources may represent plausible candidates
for the UHECR accelerators; in this case, the arrival directions would
concentrante around them.

\textbf{``New-physics'' scenarios.}

\underline{\textit{Scenario 3. Superheavy dark matter.}}
While the Large Hadron Collider failed to discover easily any dark-matter
candidate, models of dark matter which are beyond the reach of this
machine are becoming more and more popular. In particular, the superheavy
(mass $M \gtrsim 10^{18}$~eV) dark matter (SHDM) scenario, originally put
forward \cite{SHDM} to explain the apparent excess of $E \gtrsim
10^{20}$~eV cosmic rays (presently disfavoured), has its own cosmological
motivation. Its important prediction is a significant fraction of
secondary photons among the decay products of these superheavy particles;
these energetic photons contribute to the UHECR flux. For $M \gtrsim
10^{20}$~eV, the scenario is constrained, but not killed \cite{KRT-SHDM},
by the UHE photon limits; constraints for lower $M$ have not been studied.
A characteristic manifestation of this mechanism is the Galactic
anisotropy \cite{DubTin} of the arrival directions of photons related to a
non-central position of the Sun in the Galaxy.

\underline{\textit{Scenario 4. Axion-like particles and BL Lac
correlations.}}
The UHECR data set with the best angular resolution ever achieved
($0.6^{\circ}$), that of \textit{High Resolution Fly's Eye} (HiRes) in the
stereo mode, demonstrated hard-to-explain correlations of arrival
directions of $E \gtrsim 10^{19}$~eV events with distant astrophysical
sources, BL Lac type objects \cite{BLL, BLL-H}, which suggest that $\sim
2\%$ of the CR flux at these energies are neutral particles arriving from
these objects. The only self-consistent explanation of this phenomenon
\cite{axion} which does not require violation of the Lorentz invariance
suggests that the observed events are caused by the gamma rays which mix
with hypothetical new light particles (axion-like particles, ALPs) in
the cosmic magnetic fields. This would allow them to propagate freely
through the cosmic photon background in the form of the inert ALP and then
to convert back to real photons in a region of the magnetic field close to
the observer. This approach may also explain some other astrophysical
puzzles. A test of this scenario may be performed by cross-correlation of
the arrival directions with the same BL Lac catalog as in Refs.~\cite{BLL,
BLL-H}.

\underline{\textit{Scenario 5. Lorentz-invariance violation.}}
There is no lack of theoretical models with tiny violation of the
relativistic invariance on the market. In some of them, this effect
results in efficient increase of the mean free path of an energetic photon
through CMB \cite{Lorentz-violation}. Though these models have many free
parameters, with no particularly motivated choice, one may expect that a
possible effect of this change of the attenuation length would be to
increase the cosmogenic-photon flux at $E \lesssim 10^{17}$~eV by orders
of magnitude. There is no evident signature of this scenario in arrival
directions.

\subsection{Distribution of arrival directions of muonless events.}

\begin{figure}
\centering
\includegraphics[width=0.65\columnwidth]{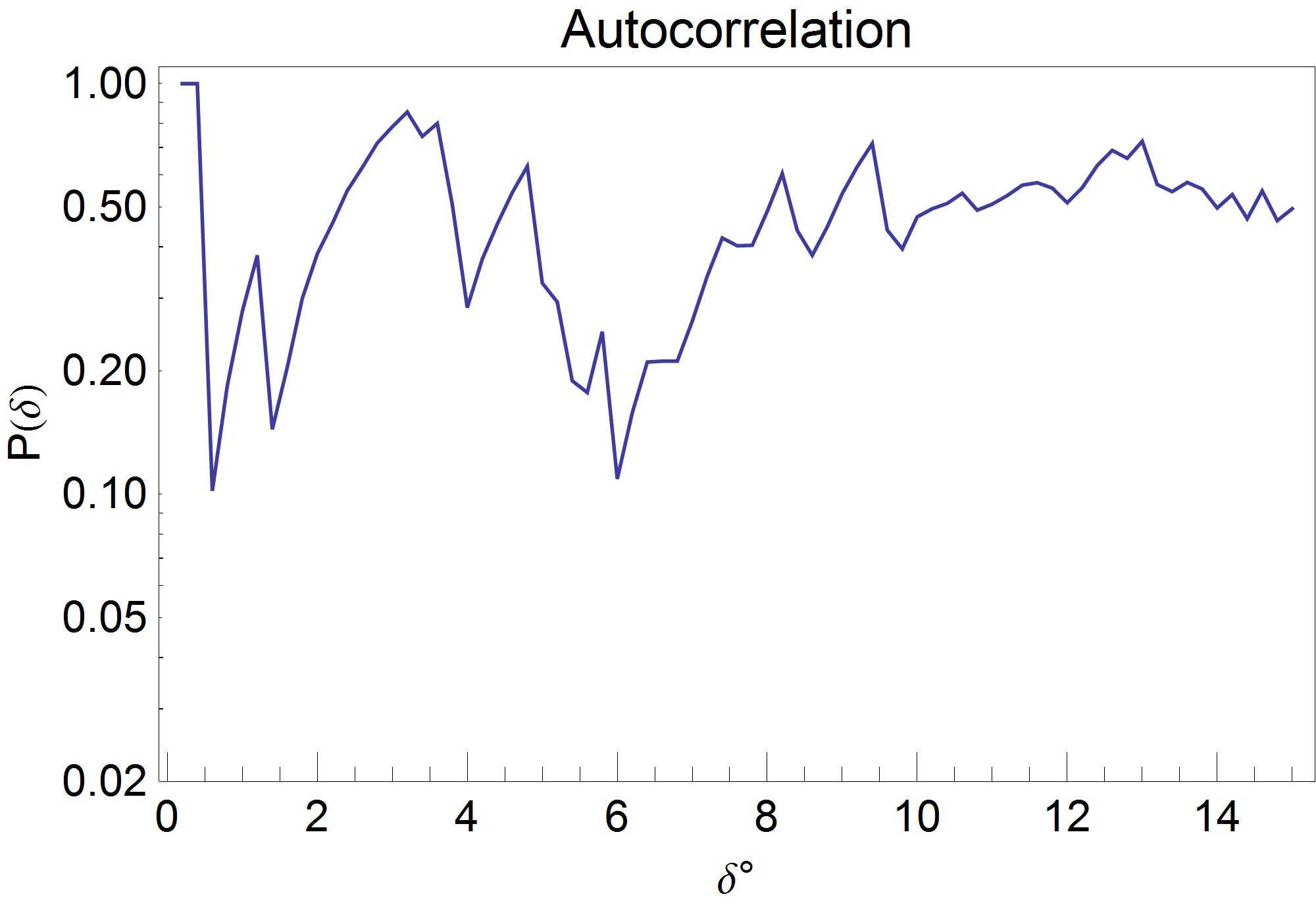}
\caption{
\label{fig:autocorr}
The autocorrelation test:
the probability $P(\Delta)$ to have the observed or higher number of pairs
of events within the angular bin $\Delta$ as a fluctuation of the
isotropic distribution.
}
\end{figure}

\textbf{1. Point sources or diffuse?}
The test of the presence of a relatively small number of point sources  is
provided by the autocorrelation function. We present the results in
Fig.~\ref{fig:autocorr}
where the probability that the observed excess of pairs of events in the
angular bin $\Delta$ is plotted as a function of $\Delta$. No significant
clustering of events is found.

\begin{figure}
\centering
\includegraphics[width=0.65\columnwidth]{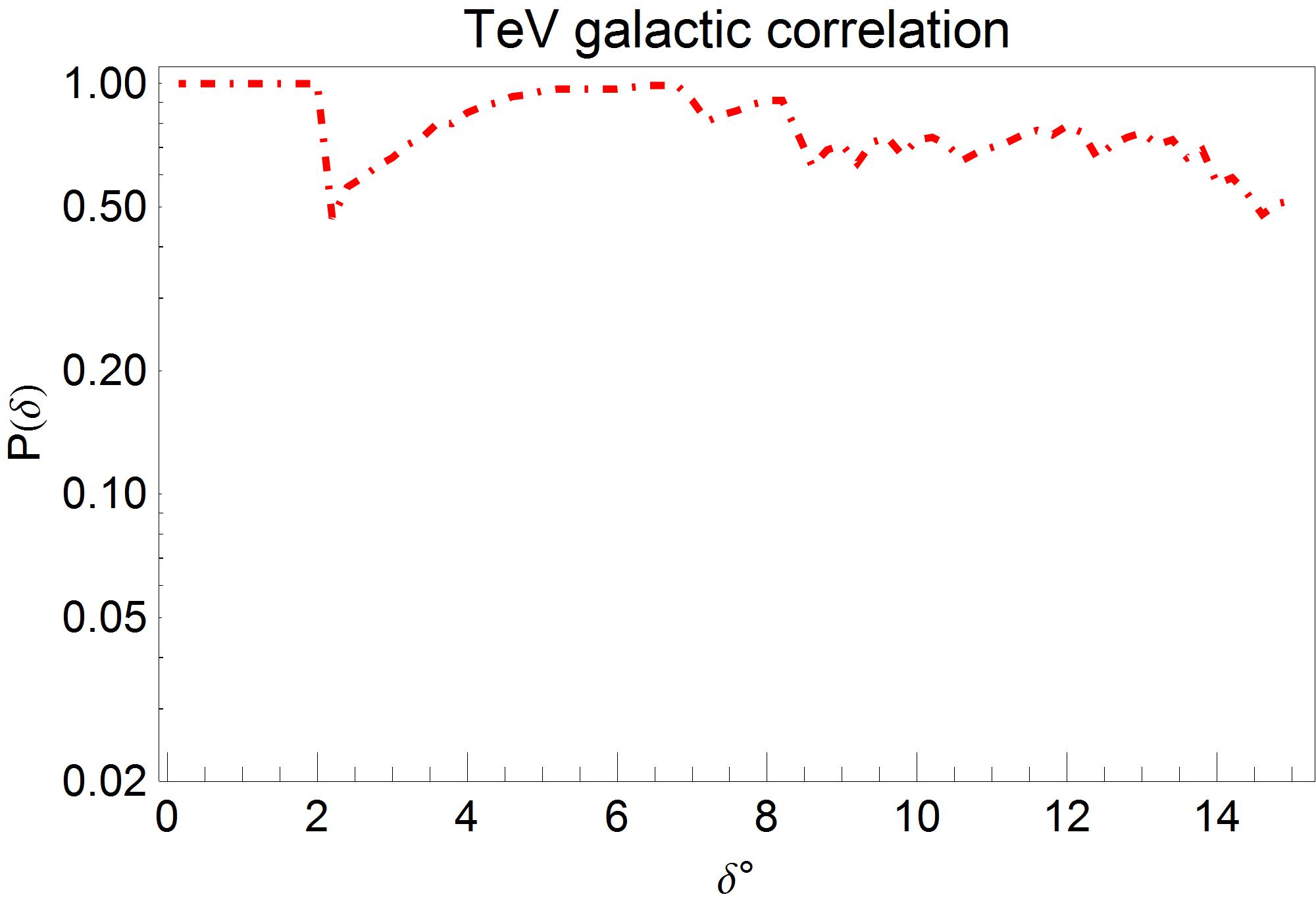}
\caption{
\label{fig:TeVGal}
The test of correlation with Galactic TeV sources:
the probability $P(\Delta)$ to have the observed or higher number
of events within the angular distance $\Delta$ from TeVCat~\cite{TeVCat}
Galactic TeV sources as a fluctuation of the isotropic distribution. }
\end{figure}

\textbf{2. Test of scenario 2: Galactic TeV sources.}
Fig.~\ref{fig:TeVGal}
represents the $P(\Delta)$ function for cross-correlations of the arrival
directions of the photon-like events with positions of Galactic TeV
sources from the TeVCat catalog~\cite{TeVCat}, as of May 2013. No sign of
correlation is seen.

\begin{figure}
\centering
\includegraphics[width=0.65\columnwidth]{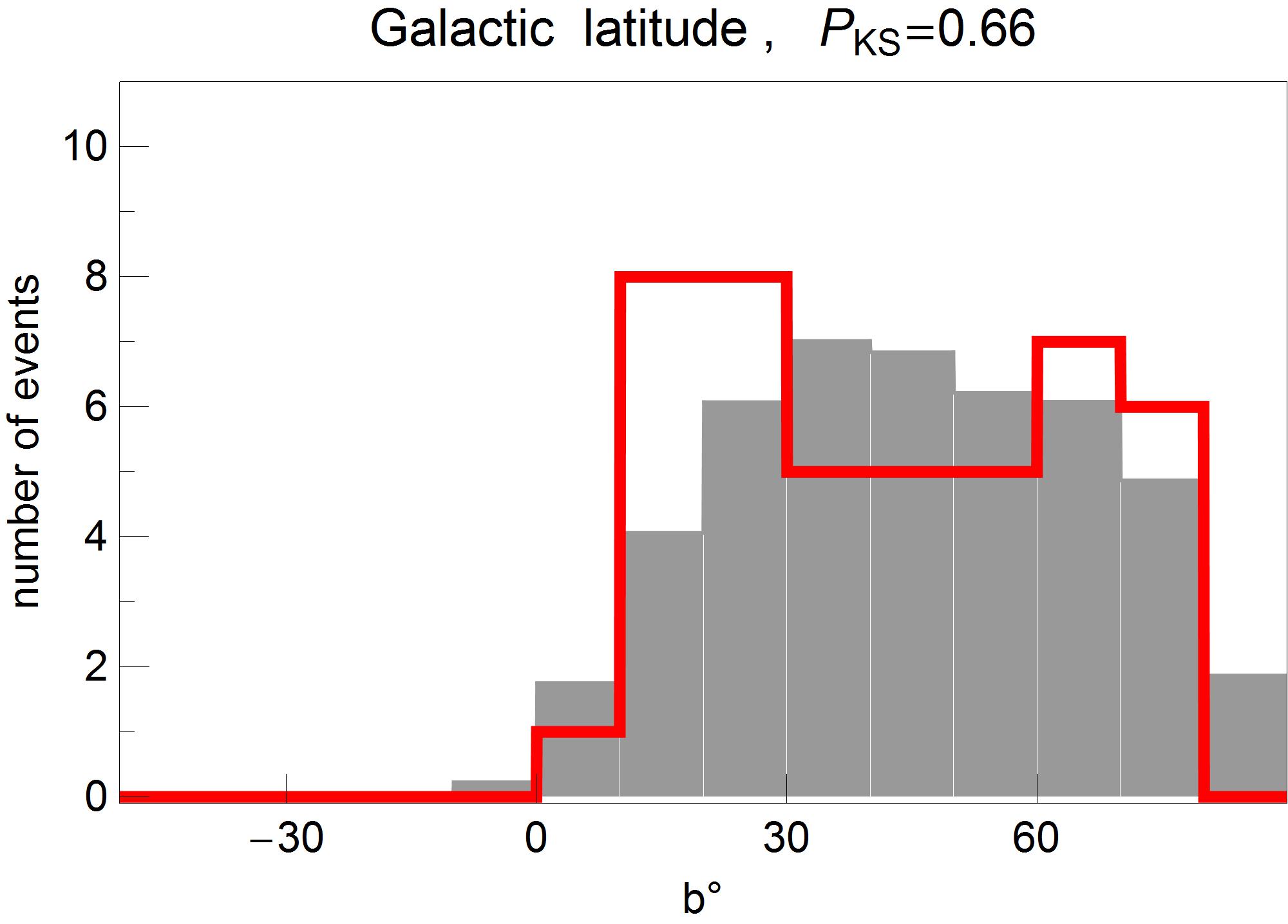}
\caption{
\label{fig:b-distr}
The distribution of the observed photon-like events (line) and Monte-Carlo
isotropic events (shadow) in the Galactic latitude $b$.}
\end{figure}

\textbf{3. Test of scenario 2: Galactic-plane correlation.}
Galactic UHECR accelerators of a yet unknown type are still expected to
concentrate along the Galactic plane, and the distribution of events in
the Galactic latitude $b$ is a model-independent test of this scenario.
Fig.~\ref{fig:b-distr}
illustrates that the distribution of the photon-like events in $b$ is
consistent with that expected for an isotropic flux (the
Kolmogorov-Smirnov probability $P_{\rm KS}\approx 0.66$).

\begin{figure}
\centering
\includegraphics[width=0.65\columnwidth]{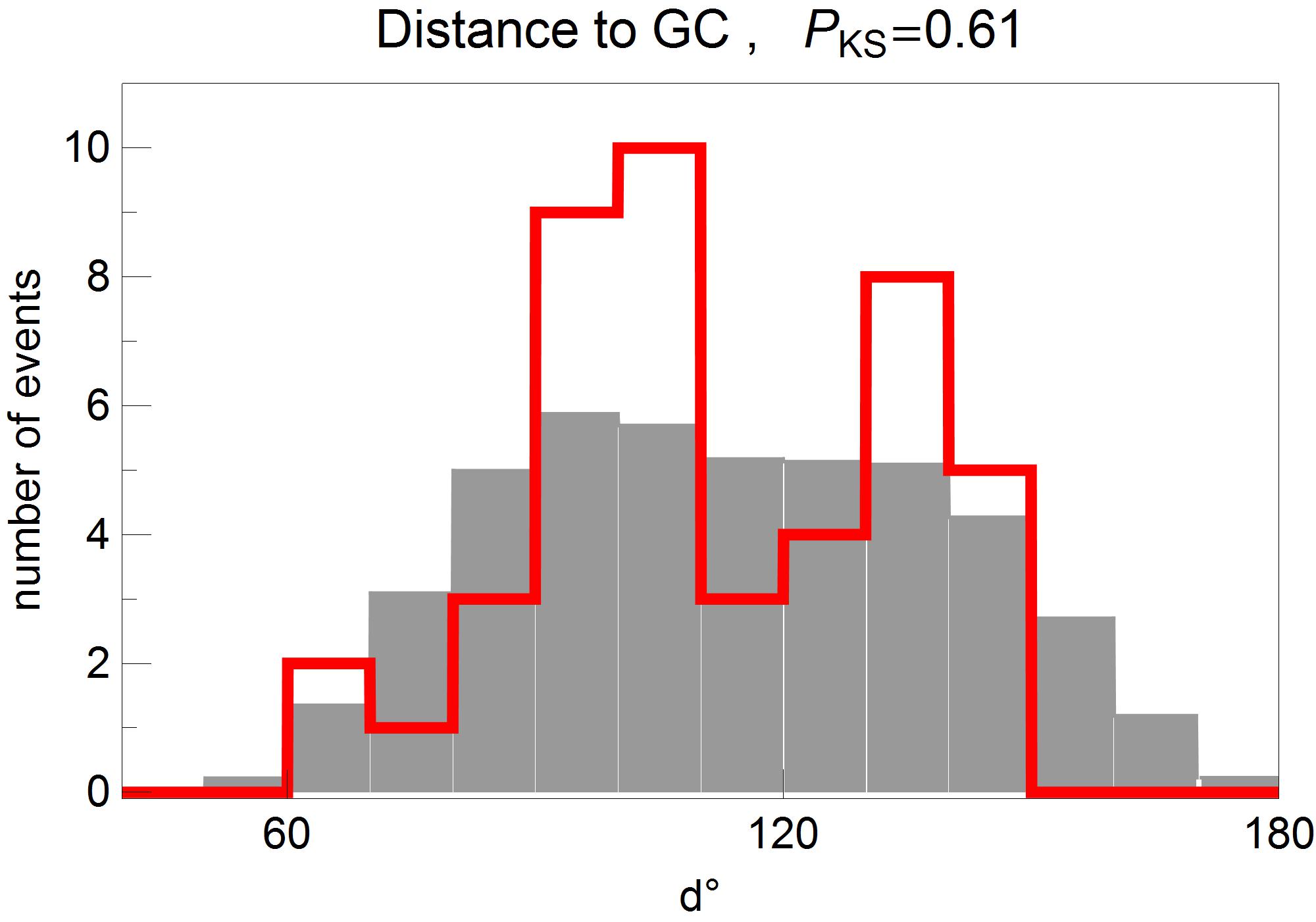}
\caption{
\label{fig:GC}
The distribution of the observed photon-like events (line) and Monte-Carlo
isotropic events (shadow) in the angular distance to the Galactic Center.}
\end{figure}

\textbf{4. Test of scenario 3: Galactic anisotropy.}
The SHDM-related Galactic anisotropy should reveal itself in the dipole
excess seen in distribution of events in the distance to the Galactic
Center. Fig.~\ref{fig:GC}
demonstrates that no such excess is seen ($P_{\rm KS}\approx 0.61$).

\begin{figure}
\centering
\includegraphics[width=0.65\columnwidth]{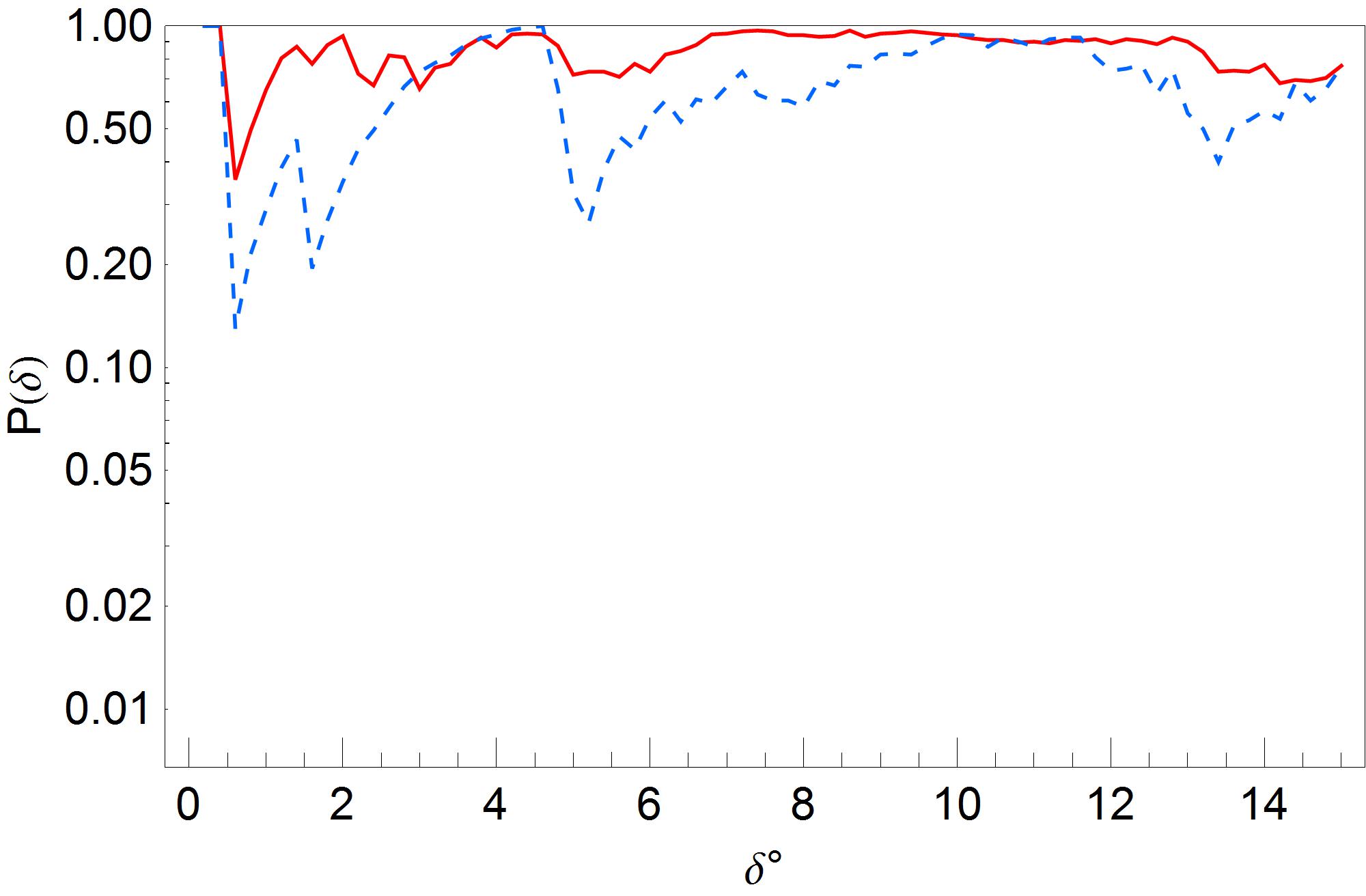}
\caption{
\label{fig:BL}
The test of correlation with BL Lac type objects:
the probability $P(\Delta)$ to have the observed or higher number
of events within the angular distance $\Delta$ from bright V\`eron BL Lacs
(sample of Ref.~\cite{BLL}, full line) and TeVCat~\cite{TeVCat} TeV BL
Lacs
(dashed line) as a fluctuation of the isotropic distribution.}
\end{figure}

\textbf{5. Test of scenario 4: BL Lac correlations.}
The HiRes BL Lac correlations~\cite{BLL} appeared as an excess of events
close to positions of 156 bright BL Lac type objects selected from the
catalog \cite{Veron} by the cut on the optical magnitude $V<18^{\rm m}$. A
subsequent study~\cite{BLL-H}  suggested also a correlation with
TeV-selected BL Lacs. In Fig.~\ref{fig:BL},
we present the results of a similar analysis for our photon-like sample,
with the same catalog of 156 BL Lacs and with an updated list of TeV BL
Lacs from TeVCat~\cite{TeVCat}. No significant correlation is seen.

\section{Discussion and conclusions}
\label{sec:concl}
\begin{figure}
\centering \includegraphics[width=0.75\columnwidth]{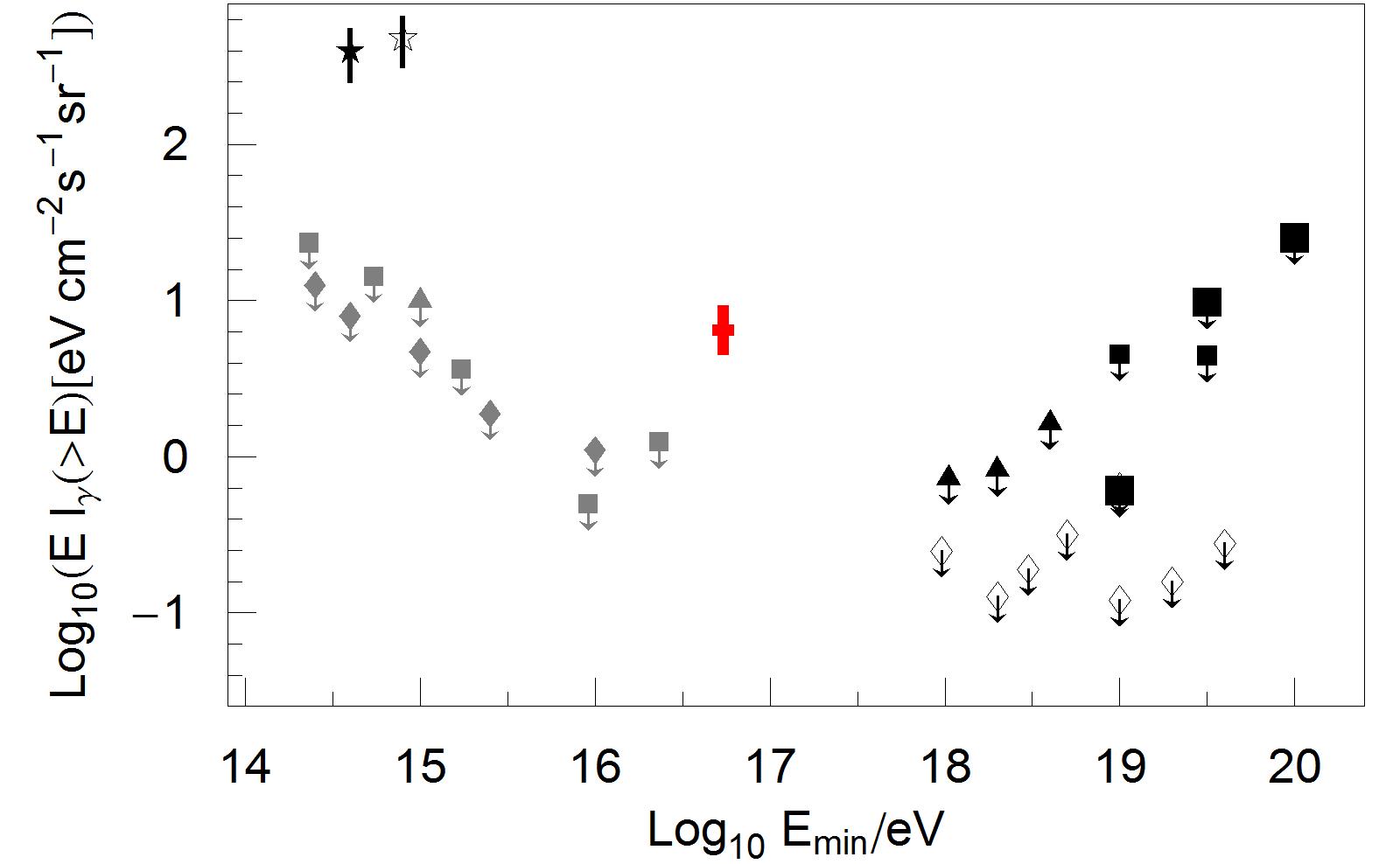}
\caption{
\label{fig:flux}
The diffuse cosmic photon integral flux versus the photon minimal
energy. The result of the present work is shown as a cross whose vertical
line represents the error bars. Tentative detections
and upper limits from other experiments are indicated by
symbols: star (Tien Shan~\cite{Tien}, detection), open star
(Lodz~\cite{Gawin}, detection), gray triangle (EAS-TOP~\cite{EAS-TOP}),
gray squares (CASA-MIA~\cite{CASA-MIA}), gray diamonds
(KASCADE~\cite{KASCADE, MMWG}), triangles (Yakutsk~\cite{Yak2}), open
diamonds (Pierre Auger \cite{PAO1, PAO2}), boxes (AGASA~\cite{AGASA}),
large squares (Telescope Array~\cite{TA}). }
\end{figure}

\begin{figure}
\centering \includegraphics[width=0.75\columnwidth]{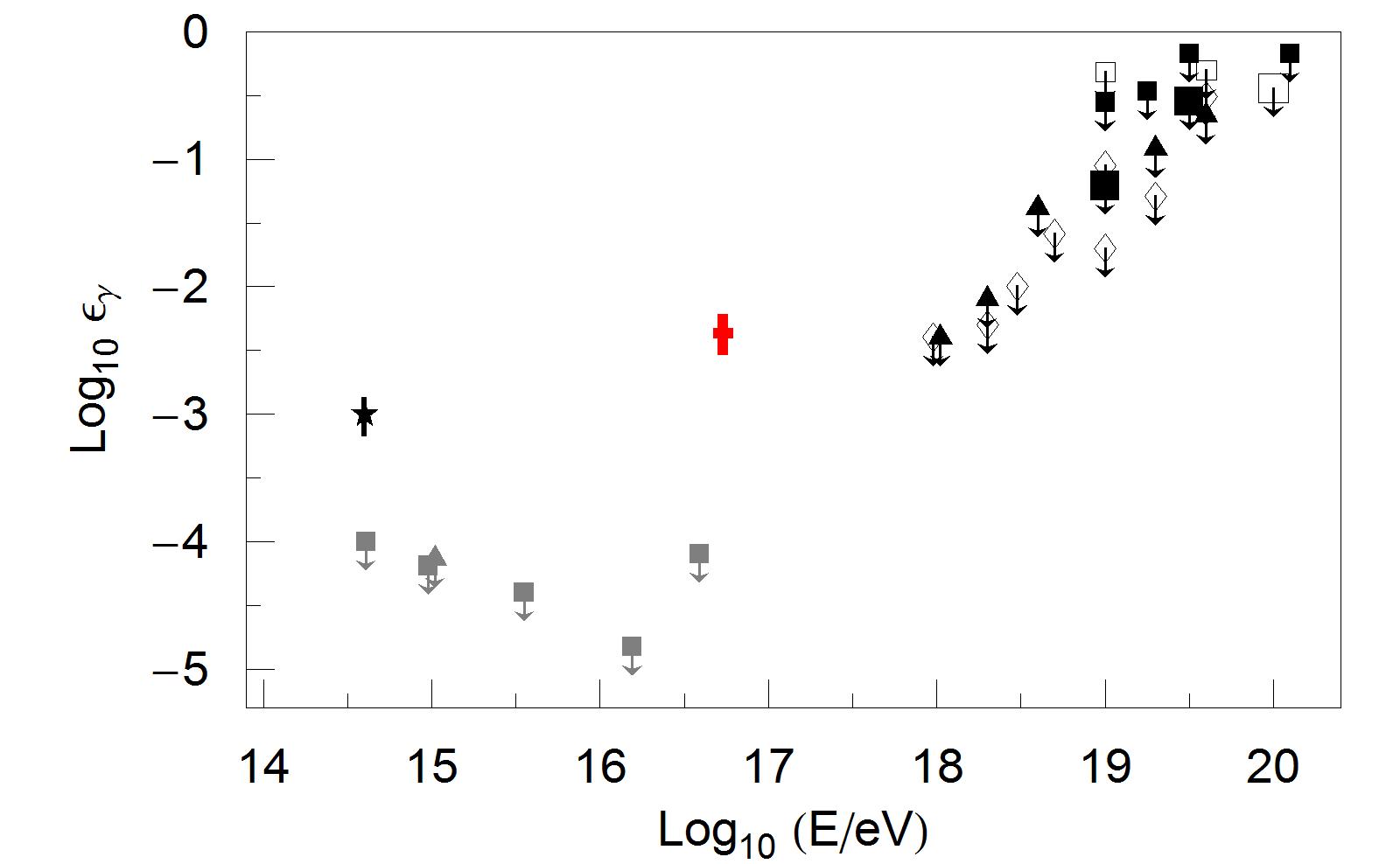}
\caption{
\label{fig:fraction}
The fraction of gamma-ray primaries in the diffuse cosmic-ray integral
flux versus the photon minimal energy.
Notations are the same as in Fig.~\ref{fig:flux}; in addition, more
Yakutsk results~\cite{Yak1}, results from Haverah Park (\cite{Haverah},
open squares), from reanalysis of the AGASA data (\cite{RisseHomola}, the
highest-energy square like AGASA) and from a combination of AGASA and
Yakutsk data (\cite{AGYak}, large open square) are shown. }
\end{figure}

The place of our result among others is rather specific. All previous
studies put upper limits on the photon flux or fraction for the primary
energy intervals $\sim (10^{14} - 5\times 10^{16})$~eV and $\gtrsim
10^{18}$~eV. The EAS-MSU result, first reported in Ref.~\cite{Khorkhe},
therefore represents the first ever statistically significant detection of
cosmic photons with energies above $\sim 100$~TeV. In the present work, we
performed the first estimate of the gamma-ray flux in the previously
unstudied energy window $(5\times 10^{16} - 10^{18})$~eV and estimated
statistical and systematic errors for its value. The result is compared
with limits obtained by other experiments in Fig.~\ref{fig:flux}
(flux) and Fig.~\ref{fig:fraction}
(fraction). The fraction estimates should be interpreted with great care
because they are sensitive to the energy determination of the bulk of
hadronic primaries which is known to suffer from large systematic
uncertainties due to lack of understanding of high-energy hadronic
interactions. Contrary, the photon flux estimates are more robust because
they use the primary gamma-ray energy determination and the exposure of
the array only, both quantities being well understood. Therefore,
the main result of this paper is the flux estimate, Eq.~(\ref{eq:flux}).

The interpretation of the result is problematic. Leaving aside the
discrepancy with the CASA-MIA result in terms of the (uncertain) gamma-ray
fraction, the more robust flux estimate, Eq.~(\ref{eq:flux}), does not
formally contradict to any existing experimental constraint but is clearly
in conflict with the general trend observed both at lower and higher
energies\footnote{We note in passing that Eq.~(\ref{eq:flux}) agrees
well with the early Yakutsk estimate~\cite{YakClaim} of
the $\sim 10^{17}$~eV photon flux, but that claim was based on observation
of one event only.}, see Fig.~\ref{fig:flux}.
Within conventional scenarios, these photons cannot travel for longer than
a few dosen Mpc and should therefore be born in the Galaxy. However, we do
not see any significant Galactic (nor any other) anisotropy in the
distribution of the arrival directions. To add to the troubles, in some
scenarios it would be difficult to avoid a conflict with measurements
of the $\sim 1$~GeV diffuse photon flux to which secondary photons from
electromagnetic cascades in the Universe have to contribute.

The estimates we made were obtained in the assumption that those muonless
events which are not accounted for by fluctuations of hadronic showers,
and only those, are caused by primary photons. This assumption is a
reasonable first approximation but it suggests two directions for the
future work. Firstly, gamma-ray showers have low but nonzero number of
muons (the reason for appearence of muons is in the photonuclear
interactions). The presence of a certain number of muonless events
implies, within the photon hypothesis, that there should be an excess of
muon-poor events in the data, which is yet to be tested. Secondly, there
are other observables, not directly related to the muon number, which may
distinguish photon showers from hadronic ones. One approach is to study
the shower front curvature which is related to the depth of the maximal
development of the electromagnetic cascade; it has been used to search for
primary photons in the experiments which do not have muon detectors,
e.g.~\cite{TA}. This method is particularly prospective for the EAS-MSU
data because the array was dense and the number of detector stations which
recorded a $N_{e} \gtrsim 10^{7}$ shower was typically large.

The result we present may be tested with the muon data of the Yakutsk EAS
array and in future experiments like fluorescence detectors of the
Telescope-Array low-energy extension (TALE) \cite{TALE}, muon detectors of
the Pierre Auger Observatory infill array (AMIGA) \cite{AMIGA} or Cerenkov
and muon detectors of the Tunka-HiSCORE \cite{HiSCORE}.

The study of arrival directions of muonless events did not
reveal any significant deviation from isotropy which might give a clue to
their origin. In principle, this may change with the extension of the data
set to an energy-limited sample (versus $N_{e}$ limited one), which would
increase the statistics for inclined events, together with more precise
determination of the arrival directions and reduction of the background by
means of two-parametric (e.g.\ both muon number and shower-front
curvature) selection of photon-like showers. We leave these questions for
a future study.

We are indebted to O.~Kalashev and G.~Rubtsov for helpful discussions. The
work of N.K., G.K., V.S.\ and Yu.F.\ was supported in part by the RFBR
grant 11-02-00544 and by the Ministry of Science and Education of the
Russian Federation, agreement 14.518.11.7046. The work of S.T.\ was
supported in part by the RFBR grants 11-02-01528, 12-02-01203,
13-02-01311, 13-02-01293 and by the Ministry of Science and Education of
the Russian Federation, agreements 8525 and 14.B37.21.0457.


\begin{thebibliography}{58}

\bibitem{KhristiansenBook}
G.B.~Khristiansen, G.V.~Kulikov, Yu.A.~Fomin,
 {\it Ultra-high-energy cosmic radiation}, Moscow, Atomizdat, 1975.

\bibitem{RisseReview}
M.~Risse and P.~Homola,
  Mod.\ Phys.\ Lett.\ A {\bf 22} (2007) 749
  [astro-ph/0702632 [ASTRO-PH]].

\bibitem{MILAGRO-90TeV}
T.~Tanimori, K.~Sakurazawa, S.A.~Dazeley {\it et al.}
[CANGAROO Collaboration],
  Astrophys.\ J.\  {\bf 492} (1998) L33
  [astro-ph/9710272].

\bibitem{EAS-TOP}
M.~Aglietta, B.~Alessandro, P.~Antoni {\it et al.}  [EAS-TOP Collaboration],
  Astropart.\ Phys.\  {\bf 6} (1996) 71.

\bibitem{CASA-MIA}
M.~C.~Chantell, C.E. Covault, J.W. Cronin {\it et al.}  [CASA-MIA
Collaboration],
  Phys.\ Rev.\ Lett.\  {\bf 79} (1997) 1805
  [astro-ph/9705246].

\bibitem{KASCADE}
G.~Schatz, F.~Fessler, T.~Antoni
{\it et al.} [KASCADE collaboration],
Proc.~28th ICRC, Tsukuba {\bf 4} (2003) 2293

\bibitem{Haverah}
M.~Ave, J.~A.~Hinton, R.~A.~Vazquez   {\it et al.}
  Phys.\ Rev.\ Lett.\  {\bf 85} (2000) 2244
  [astro-ph/0007386].

\bibitem{AGASA}
K.~Shinozaki, M.~Chikawa, M.~Fukushima {\it et al.} [AGASA Collaboration],
  Astrophys.\ J.\  {\bf 571} (2002) L117.

\bibitem{RisseHomola}
M.~Risse, P.~Homola, R.~Engel {\it et al.},
  Phys.\ Rev.\ Lett.\  {\bf 95} (2005) 171102
  [astro-ph/0502418].

\bibitem{AGYak}
G.~I.~Rubtsov, L.~G.~Dedenko, G.~F.~Fedorova  {\it et al.},
  Phys.\ Rev.\ D {\bf 73} (2006) 063009
  [astro-ph/0601449].

\bibitem{Yak1}
 A.~V.~Glushkov, D.~S.~Gorbunov, I.~T.~Makarov {\it et al.},
  JETP Lett.\  {\bf 85} (2007) 131
  [astro-ph/0701245].

\bibitem{Yak2}
A.~V.~Glushkov, I.~T.~Makarov, M.~I.~Pravdin  {\it et al.},
  Phys.\ Rev.\ D {\bf 82} (2010) 041101
  [arXiv:0907.0374 [astro-ph.HE]].

\bibitem{PAO1}
J.~Abraham, P.~Abreu, M.~Aglietta {\it et al.}
[Pierre Auger Collaboration],
  Astropart.\ Phys.\  {\bf 29} (2008) 243
  [arXiv:0712.1147 [astro-ph]].

\bibitem{PAO2}
P.~Abreu, M.~Aglietta, E.J.~Ahn  {\it et al.}
[Pierre Auger Collaboration],
  arXiv:1107.4805 [astro-ph.HE].

\bibitem{TA}
T.~Abu-Zayyad, R.~Aida, M.~Allen {\it et al.}
[Telescope Array Collaboration],
  arXiv:1304.5614 [astro-ph.HE].

\bibitem{Chac}
K.~Suga, Y.~Toyoda, K.~Kamata   {\it et al.},
  Astrophys.\ J.\  {\bf 326} (1988) 1036.

\bibitem{Tien}
S.~I.~Nikolsky, I.~N.~Stamenov and S.~Z.~Ushev,
  J.\ Phys.\ G {\bf 13} (1987) 883.

\bibitem{YakClaim}
A.V.~Glushkov, N.N.~Efimov, N.N.~Efremov {\it et al.},
Proc.\ 19th ICRC, La Jolla, \textbf{2} (1985) 186.

\bibitem{Gawin}
J.~Gawin, R.~Maze, J.~Wdowczyk and A.~Zawadzki,
Canad.\ J.\ Phys.\ {\bf 46} (1968) 75.

\bibitem{Unger}
K.~-H.~Kampert and M.~Unger,
  Astropart.\ Phys.\  {\bf 35} (2012) 660
  [arXiv:1201.0018 [astro-ph.HE]].

\bibitem{ST-UFN}
S.~Troitsky,
  Phys.\  Usp.\  {\bf 56} (3) (2013), Uspekhi Fiz. Nauk {\bf 183} (2013) 323
  [arXiv:1301.2118 [astro-ph.HE]].

\bibitem{BhSigl}
P.~Bhattacharjee and G.~Sigl,
  Phys.\ Rept.\  {\bf 327} (2000) 109
  [astro-ph/9811011].

\bibitem{Lorentz-violation}
 M.~Galaverni and G.~Sigl,
  Phys.\ Rev.\ Lett.\  {\bf 100} (2008) 021102
  [arXiv:0708.1737 [astro-ph]].

\bibitem{axion}
M.~Fairbairn, T.~Rashba and S.~V.~Troitsky,
  Phys.\ Rev.\ D {\bf 84} (2011) 125019
  [arXiv:0901.4085 [astro-ph.HE]].

\bibitem{EAS-MSU}
S.~N.~Vernov, G.~B.~Khristiansen, V.~B.~Atrashkevich
{\it et al.},
  Bull.\ Russ.\ Acad.\ Sci.\ Phys.\  {\bf 44} (1980) 80
   [Izv.\ Ross.\ Akad.\ Nauk Ser.\ Fiz.\  {\bf 44} (1980) 537].

\bibitem{Khorkhe}
 N.~Kalmykov, J.~Cotzomi, V.~Sulakov   {\it et al.},
Izv.\ Ross.\ Akad.\ Nauk Ser.\ Fiz.\  {\bf 75} (2009) 584

\bibitem{muLDF}
Yu.A.~Fomin, N.N.~Kalmykov, V.M.~Kalmykov   {\it et al.},
Proc. 28th ICRC, Tsukuba, {\bf 1} (2003) 119

\bibitem{AIRES}
 S.J.~Sciutto, {\it AIRES: A system for air shower simulations. Version
2.6.0}, 2002.

\bibitem{QGSJET}
N.~N.~Kalmykov, S.~S.~Ostapchenko and A.~I.~Pavlov,
  Nucl.\ Phys.\ Proc.\ Suppl.\  {\bf 52B} (1997) 17.

\bibitem{assumption}
A.~A.~Lagutin, V.~V.~Uchaikin and G.~V.~Chernyaev,
  Yad.\ Fiz.\  {\bf 45} (1987) 757.

\bibitem{EAS-MSUflux}
N.N.~Kalmykov, L.A.~Kuzmichev, G.V.~Kulikov
{\it et al.},
Moscow Univ.\ Phys.\ Bull.\ {\bf 65} (2010) 275.

\bibitem{Q01-mult}
D.~d'Enterria, R.~Engel, T.~Pierog     {\it et al.},
  Astropart.\ Phys.\  {\bf 35} (2011) 98
  [arXiv:1101.5596 [astro-ph.HE]].

\bibitem{EngelTalkUHECR}
R.~Engel, Talk at the
International Symposium on Future Directions in UHECR Physics, CERN,
13-16 February 2012.

\bibitem{SIBYLL}
E.~-J.~Ahn, R.~Engel, T.~K.~Gaisser   {\it et al.},
  Phys.\ Rev.\ D {\bf 80} (2009) 094003
  [arXiv:0906.4113 [hep-ph]].

\bibitem{QGSJET2}
S.~Ostapchenko,
  Nucl.\ Phys.\ Proc.\ Suppl.\  {\bf 151} (2006) 143
  [hep-ph/0412332].

\bibitem{EPOS}
 T.~Pierog and K.~Werner,
  Nucl.\ Phys.\ Proc.\ Suppl.\  {\bf 196} (2009) 102
  [arXiv:0905.1198 [hep-ph]].

\bibitem{PAOmuons}
 R.~Engel [Pierre Auger Collaboration],
  arXiv:0706.1921 [astro-ph].

\bibitem{Yakutsk-mu}
 A.~V.~Glushkov, I.~T.~Makarov, M.~I.~Pravdin  {\it et al.},
  JETP Lett.\  {\bf 87} (2008) 190
  [arXiv:0710.5508 [astro-ph]].

\bibitem{Yakutsk-mu1}
L.~G.~Dedenko, G.~F.~Fedorova, T.~M.~Roganova
{\it et al.},
  J.\ Phys.\ G {\bf 39} (2012) 095202.

\bibitem{Uchaikin}
V.V.~Uchaikin, V.V.~Ryzhov,
{\it The Stochastic Theory of High Energy Particle Transport},
Novosibirsk, Nauka, Siberian Branch (1988) (in Russian).

\bibitem{Khorkhe-diss}
J.~Cotzomi Paleta, PhD Thesis, SINP MSU, Moscow, 2010.

\bibitem{nr}
W.~H.~Press, S.~A.~Teukolsky, W.~T.~Vetterling and B.~P.~Flannery,
{\it Numerical Recipes: The Art of Scientific Computing},
 Cambridge University Press, 2007.

\bibitem{cut-n-pen}
P.~Tinyakov and I.~Tkachev,
  Phys.\ Rev.\ D {\bf 69} (2004) 128301
  [astro-ph/0301336].

\bibitem{G}
K.~Greisen,
  Phys.\ Rev.\ Lett.\  {\bf 16} (1966) 748.

\bibitem{ZK}
 G.~T.~Zatsepin and V.~A.~Kuzmin,
  JETP Lett.\  {\bf 4} (1966) 78
   [Pisma Zh.\ Eksp.\ Teor.\ Fiz.\  {\bf 4} (1966) 114].

\bibitem{Gelmini}
G.~Gelmini, O.~E.~Kalashev and D.~V.~Semikoz,
  J.\ Exp.\ Theor.\ Phys.\  {\bf 106} (2008) 1061
  [astro-ph/0506128].

\bibitem{Sarkar}
D.~Hooper, A.~M.~Taylor and S.~Sarkar,
  Astropart.\ Phys.\  {\bf 34} (2011) 340
  [arXiv:1007.1306 [astro-ph.HE]].

\bibitem{SHDM}
V.~Berezinsky, M.~Kachelriess and A.~Vilenkin,
  Phys.\ Rev.\ Lett.\  {\bf 79} (1997) 4302
  [astro-ph/9708217].

\bibitem{KRT-SHDM}
O.~E.~Kalashev, G.~I.~Rubtsov and S.~V.~Troitsky,
  Phys.\ Rev.\ D {\bf 80} (2009) 103006
  [arXiv:0812.1020 [astro-ph]].

\bibitem{DubTin}
S.~L.~Dubovsky, P.~G.~Tinyakov and I.~I.~Tkachev,
  Phys.\ Rev.\ Lett.\  {\bf 85} (2000) 1154
  [astro-ph/0001317].

\bibitem{BLL}
D.~S.~Gorbunov, P.~G.~Tinyakov, I.~I.~Tkachev and S.~V.~Troitsky,
  JETP Lett.\  {\bf 80} (2004) 145
   [Pisma Zh.\ Eksp.\ Teor.\ Fiz.\  {\bf 80} (2004) 167]
  [astro-ph/0406654].

\bibitem{BLL-H}
R.~U.~Abbasi {\it et al.}  [HiRes Collaboration],
  Astrophys.\ J.\  {\bf 636} (2006) 680
  [astro-ph/0507120].

\bibitem{TeVCat}
TeVCat, {\tt http://tevcat.uchicago.edu/} .

\bibitem{Veron}
M.~P.~V\'eron-Cetty and P.~V\'eron,
Astron.\ Astrophys.\ {\bf 374} (2001) 92.

\bibitem{MMWG}
M.~Risse and G.~Rubtsov, Talk at the
International Symposium on Future Directions in UHECR Physics, CERN,
13-16 February 2012.

\bibitem{TALE}
S.~Ogio {\it et al.} [Telescope Array Collaboration],
Talk at the
International Symposium on Future Directions in UHECR Physics, CERN,
13-16 February 2012.

\bibitem{AMIGA}
A.~Etchegoyen [Pierre Auger Collaboration],
  arXiv:0710.1646 [astro-ph].

\bibitem{HiSCORE}
 M.~Tluczykont, D.~Hampf, U.~Einhaus
{\it et al.},
  AIP Conf.\ Proc.\  {\bf 1505} (2012) 821.

\end{thebibliography}
\end{document}